\documentclass{elsarticle}

\usepackage[utf8]{inputenc}
\usepackage{graphicx}
\usepackage{mathtools}
\usepackage{cleveref}
\usepackage{siunitx}
\usepackage[nolist]{acronym}
\usepackage{todonotes}
\usepackage{subcaption}
\usepackage{booktabs}
\usepackage{float}
\usepackage{listings}
\usepackage{fancybox}
\usepackage{xcolor}
\usepackage{extdash}
\usepackage{filecontents}

\colorlet{punct}{red!60!black}
\definecolor{delim}{RGB}{20,105,176}
\definecolor{keyword}{RGB}{15,95,210}
\colorlet{numb}{magenta!100!black}

\lstdefinelanguage{json}{
    basicstyle=\normalfont\ttfamily,
    showstringspaces=false,
    breaklines=true,
    frame=none,
    literate=
     *{0}{{{\color{numb}0}}}{1}
      {1}{{{\color{numb}1}}}{1}
      {2}{{{\color{numb}2}}}{1}
      {3}{{{\color{numb}3}}}{1}
      {4}{{{\color{numb}4}}}{1}
      {5}{{{\color{numb}5}}}{1}
      {6}{{{\color{numb}6}}}{1}
      {7}{{{\color{numb}7}}}{1}
      {8}{{{\color{numb}8}}}{1}
      {9}{{{\color{numb}9}}}{1}
      {true}{{{\color{keyword}true}}}{4}
      {false}{{{\color{keyword}false}}}{4}
      {null}{{{\color{keyword}null}}}{4}
      {:}{{{\color{punct}{:}}}}{1}
      {,}{{{\color{punct}{,}}}}{1}
      {\{}{{{\color{delim}{\{}}}}{1}
      {\}}{{{\color{delim}{\}}}}}{1}
      {[}{{{\color{delim}{[}}}}{1}
      {]}{{{\color{delim}{]}}}}{1},
}

\makeatletter
\newenvironment{CenteredBox}{%
\begin{Sbox}}{
\end{Sbox}\centerline{\parbox{\wd\@Sbox}{\TheSbox}}}
\makeatother

\acrodefplural{POI}{points of interest}

\begin{document}
\begin{frontmatter}
\title{OMOD: An open-source tool for creating disaggregated mobility demand based on OpenStreetMap}
\author[1]{Leo Strobel\corref{cor1}}
\ead{leo.strobel@uni-wuerzburg.de}
\author[1]{Marco Pruckner}
\ead{marco.pruckner@uni-wuerzburg.de}
\address[1]{University of Würzburg, Am Hubland, 97074, Würzburg, Germany}
\cortext[cor1]{Corresponding author}

\begin{abstract}
This paper introduces \acsu{OMOD} (\acl{OMOD}),
a new open-source activity-based mobility demand generation tool.
\ac{OMOD} uses a data-driven approach, calibrated with household travel survey data,
to generate a population of agents
with detailed daily activity schedules that state what activities each agent plans to conduct, where, and for how long.
The temporal aspect of the output is wholly disaggregated,
while the spatial aspect is given on the level of individual buildings.
In contrast to other existing models,
\ac{OMOD} is freely available,
open-source,
works out-of-the-box,
can be applied to anywhere in Germany with the ambition to widen the scope to other countries,
and only requires freely available \ac{OSM} data from the user.
With \ac{OMOD},
it is easy for non-experts to create realistic mobility demand,
which can be used in transportation studies, energy system modeling, communications system research, et cetera.
This paper describes \ac{OMOD}'s architecture and validates the model for three cities ranging from 200,000 to 2.5 million inhabitants.
\end{abstract}


\end{frontmatter}

\section{Introduction}
Models of human mobility are traditionally used by transportation research\Hyphdash ers to design efficient transport systems
\cite{songWholeDayPath2021,debokPopulationSimulatorDisaggregate2015,zhouSustainableMobilityStrategies2023,liuNetworkorientedHouseholdActivity2018,nguyenUnifiedActivitybasedFramework2022,iacobucciDemandPotentialShared2023}.
However, such models are useful beyond transportation research and are increasingly common in other fields like
homeland security policy research \cite{hensonAssessmentActivitybasedModeling2009},
epidemiology \cite{mahdizadehgharakhanlouSpatiotemporalSimulationNovel2020},
or communication systems research.
In communication systems research,
models that include human mobility patterns are utilized to test and optimize networking schemes
for ad hoc networks \cite{johnsonDynamicSourceRouting1996},
device-to-device communication \cite{seufertPotentialTrafficSavings2018, zhouMobileDevicetoDeviceVideo2016},
or vehicle-to-vehicle communication \cite{niebischCoDiPyPerformanceEvaluation2022}.
To this end, random movement models are often used \cite{johnsonDynamicSourceRouting1996,zhouMobileDevicetoDeviceVideo2016,niebischCoDiPyPerformanceEvaluation2022}.
However, these do not accurately depict the cyclic \cite{gonzalezUnderstandingIndividualHuman2008}
and highly predictive \cite{songLimitsPredictabilityHuman2010} aspects of human mobility behavior and,
consequently, might come to false conclusions about real-world performance.

Another field where human mobility models become increasingly relevant is energy system modeling.
In light of the tremendous numbers of electric vehicles projected to be on the streets in a decade \cite{GlobalElectricVehicle2022},
grid operators face the difficult task of ensuring grid stability.
To adequately design grid expansions,
it is necessary to understand the newly emerging electricity demand of electric vehicles.
This demand is determined by when and where electric vehicles charge.
Since the vehicles will move with their owner,
accurate models of mobility behavior are necessary \cite{liElectricVehicleBehavior2023,strobelJointAnalysisRegional2022,Iacobucci23}.

With this increasing interest in mobility behavior,
it becomes essential to have mobility demand models that are applicable in a wide variety of research fields,
can be applied to a wide range of locations,
are easy to use,
and, most importantly, do not require access to proprietary data from the end user.
A model fulfilling these conditions allows researchers in other fields to generate realistic mobility demand for their studies
without requiring knowledge of transportation modeling and access to data that is inherently connected with privacy concerns.
On the other hand,
this exchange allows modeling approaches developed in the transportation field to be tested under a changed point of view and
with different performance metrics,
which invariably leads to new insights.

This paper introduces the \acf{OMOD}, an open-source\footnote{Available on GitHub https://github.com/L-Strobel/omod under the MIT license.} activity-based simulation tool.
The main contribution of \ac{OMOD} is that it fulfills all the state requirements for interdisciplinary use described above
by being applicable to anywhere in Germany,
being open-source, useable out-of-the-box,
and not requiring any proprietary data on the user side.
\ac{OMOD} is designed such that a rapid application to other countries is possible
when a detailed household travel survey is available, and the location is sufficiently mapped in \ac{OSM}.
Furthermore, \ac{OMOD} adheres to the best practices learned in transportation research
and, therefore,
significantly improves upon the models currently used in fields like communication systems research or energy system modeling.

Under mobility demand generation,
we understand the steps of population and activity generation (similar to trip generation and distribution steps in four-step models).
Mode and route choice are left undetermined for other software like SUMO \cite{lopezMicroscopicTrafficSimulation2018} or MATSim \cite{horniMultiAgentTransportSimulation2016}. 
Therefore, \ac{OMOD} determines what a person would like to do on a given day or week
if they had the necessary means of transportation.

This paper is structured as follows.
First,
we review related work in mobility demand modeling,
focusing on open-source tools (see \Cref{sec:RelatedWork}).
Then,
we describe the architecture of \ac{OMOD} (see \Cref{sec:Architecture}).
The calibration process we applied to arrive at the default parameterization
is described alongside the model's architecture.
Finally,
we validate the model by comparing its output to the German national household travel survey \ac{MiD} \cite{infasMobilitatDeutschland2017} (see \Cref{sec:validation}).

\section{Related Work}
\label{sec:RelatedWork}

In transportation research, travel models are typically created with a specific region in mind.
Examples include,
TASHA \cite{millerPrototypeModelHousehold2003,roordaValidationTASHA24h2008}, an activity generation and scheduling model for the Greater Toronto Area,
SACISM \cite{bradleySACSIMAppliedActivitybased2010}, a model developed for the Sacramento area,
FAMOS \cite{pendyalaFloridaActivityMobility2005}, an activity-based travel demand forecasting system for the State of Florida,
or the official transport demand model for Flanders used by the Flemish Authorities \cite{debokPopulationSimulatorDisaggregate2015}.
Many more examples of these models (primarily focusing on the US) can be found in the review by Davidson et al. \cite{davidsonSynthesisFirstPractices2007}.
Here, the new generation of activity-based models is compared to conventional four-step models.
They describe a persistent gap between recent research focusing on activity-based models on the one hand and
practitioners relying on conventional models on the other hand.
Additionally, they show several examples where more modern approaches have been implemented successfully in practice.
They highlight three features that the new generation of models has, which are also present in \ac{OMOD}:

\begin{itemize}
    \item activity-based: The models derive mobility demand from the desire of each person to conduct daily activities
    (instead of directly determining trip numbers by extrapolating surveys).
    \item tour-based: The models use tours\footnote{A tour is a sequence of trips starting and ending at the same location \cite{ortuzarModellingTransport4th2011}.}
    as the basic unit of travel demand.
    Using tours ensures that trips are self-consistent,
    i.e., every trip leaving home must eventually lead to a trip returning to the home location.
    \ac{OMOD} goes one step further by using daily activity schedules,
    meaning that an individual's entire day must be consistent.
    \item micro-simulation: The mobility demand is modeled on the fully-disaggregate level of persons and households.
\end{itemize}

Shiftan and Ben-Akiva \cite{shiftanPracticalPolicysensitiveActivitybased2011} conduct a similar analysis to Davidson et al. \cite{davidsonSynthesisFirstPractices2007}.
They determine best practices that can be learned from the "best" practical activity-based models,
particularly regarding the trade-off between realism and model complexity.
Among other things, they find that the analyzed models generally:
model interactions across tours
(i.e., use day patterns or daily activity schedules to model an entire day consistently),
disregard household interactions in favor of the simplicity of independent individuals,
and determine trip destinations sequentially,
using a random subsample of all traffic assignment zones as choice set.

These models aim to create policy-sensitive forecasting tools for the mobility demand in specific regions.
Significant work is put into fine-tuning the model.
Consequently, these models work well for their purpose, but applying them to other regions is often difficult, especially for non-experts.
Additionally, in many cases,
the models are not open-source and rely on private data.
Notable exceptions to the issue of limited transferability
are the activity-travel pattern simulator CEMDAP \cite{bhatComprehensiveEconometricMicrosimulator2004}
and the activity-based transport demand modeling framework FEATHERS \cite{bellemansImplementationFrameworkDevelopment2010}.
In both cases, the software architectures have been constructed with transferability in mind.
However, both models are still geared towards transportation researchers and require the user to collect and format various data sets for the area of interest,
which might or might not be publicly available.
For example, CEMDAP requires the user to provide zone-to-zone transportation system level-of-service characteristics by time of day.
As stated in the introduction, 
more and more use cases for mobility demand models can be found in other research fields where no transportation experts and
access to proprietary data exist.
\ac{OMOD} tries to provide a model for these use cases
while simultaneously adhering to the best practices learned in the field of transportation study.

The obstacle that proprietary data represents for the transferability and verifiability of transport models
is well-known in transportation research.
Various authors have proposed synthetic population creation pipelines that rely only on public data.
Notable examples are the work of Agriesti et al. \cite{agriestiAssignmentSyntheticPopulation2022},
Felbermair et al. \cite{felbermairGeneratingSyntheticPopulation2020},
and Hörl and Balac \cite{horlSyntheticPopulationTravel2021}.
In each case,
the authors demonstrate good results for a specific case study (Tallinn, Carinthia in Austria, and Paris).
However, they all use a wide range of public data sets,
complicating the transfer of their approaches
because a substitution has to be found for each used data set,
which might be difficult or impossible depending on the area of interest.
For example, they all rely on detailed commuting origin-destination matrices.
\ac{OMOD}, on the other hand, requires only \ac{OSM} data (available for almost anywhere)
when applied to a region where a suitable calibration exists (i.e., anywhere in Germany).
Household travel survey data is necessary to calibrate \ac{OMOD} to new regions.
However, since the calibration process does not need to store any survey records (as opposed to \cite{horlSyntheticPopulationTravel2021}),
sharing calibrations is less problematic, even for non-public surveys.

Other existing models set out to provide mobility demand models that can be applied to any area.
Isaacman et al. \cite{isaacmanHumanMobilityModeling2012} introduce the WHERE model
that simulates movement patterns of individuals in a given region and
outputs them in the form of synthetic Call Detail Records.
Their work is extended by Darakhshan et al. \cite{mirDPWHEREDifferentiallyPrivate2013},
who add noise to the output to improve privacy,
and by Smolak et al. \cite{smolakPopulationMobilityModelling2020b},
who enhance the temporal component of WHERE and restrict the movement of agents to an elliptic \textit{activity space}
that encompasses their home and work location.
These models require detailed movement trajectory data as input
in the form of Call Detail Record traces
or synthesized from census data.
They rely heavily on the quality of that data, as they do not include GIS data from the modeled area.

The transportation simulators TRANSSIMS \cite{leeApplicationsTRANSIMSTransportation2014} and SUMO \cite{lopezMicroscopicTrafficSimulation2018} (called \textit{activitygen})
include basic mobility demand generators.
The former expects a household travel survey as input, the latter only common census data.
Furthermore, tools have been developed in the SUMO community to improve upon \textit{activitygen} \cite{schweizerGeneratingActivityBased,codecaSAGAActivitybasedMultimodal2020}.
Of these, especially noteworthy is SAGA \cite{codecaSAGAActivitybasedMultimodal2020}.
Like \ac{OMOD}, SAGA requires the user only to provide an \ac{OSM} file as input.
The destination choice process is similar to that in \cite{smolakPopulationMobilityModelling2020b}.
Secondary locations are restricted to an elliptic area defined by the home, primary location, and user-defined radius. 
Home and primary locations are sampled independently based on the weighted probability of each traffic assignment zone,
where the weight is the number of buildings, \acp{POI}, and infrastructure objects (i.e., streets) divided by the area of the traffic assignment zone.
While SAGA can be used solely with an \ac{OSM} file,
proper usage requires significant parameterization by the user.
Most notably, the user must provide which activity chains are possible,
with what probability they occur,
and for how long each activity lasts.
SAGA does not offer a set of precalibrated default values like \ac{OMOD}.
For this reason,
it is more suited as a tool that reduces overhead for experts rather than a tool that can be used out-of-the-box.

In energy system modeling,
several open-source models of electric vehicle mobility exist \cite{schlundElectricVehicleCharging2021,wulffVehicleEnergyConsumption2021,gaete-moralesOpenToolCreating2021a}.
The most prominent among them is \textit{emobpy} by Gaete-Morales et al. \cite{gaete-moralesOpenToolCreating2021a}.
These models are built upon household travel surveys,
but only \textit{vencopy} \cite{wulffVehicleEnergyConsumption2021} requires the user to have access to the survey itself.
The other two models \cite{schlundElectricVehicleCharging2021,gaete-moralesOpenToolCreating2021a}
ship the underlying probability distributions of the survey with the model.
Therefore, the user does not have to provide any input data.
Although the primary output of these models is the electric demand of a fleet of electric vehicles,
they work by creating trip chains for a synthetic population of agents,
and it is possible to obtain these chains directly.
However, none of these models include a destination choice model.
Instead, locations are usually only described abstractly,
like \textit{home} or \textit{work},
without providing actual coordinates,
limiting their usefulness for purposes other than electric vehicle demand.
Additionally, it hinders them from being used to analyze local electric grid congestion effects, even in energy system modeling.

Currently, no mobility demand simulator exists that adheres to the best practices of activity-based modeling,
works out-of-the-box without requiring expert knowledge or proprietary data on the user side,
and outputs fully disaggregated and spatially referenced mobility profiles.
With \ac{OMOD}, we plan to provide exactly that,
hoping the model will find users in various fields.
\ac{OMOD} aims to achieve this goal by relying on the \ac{OSM} ecosystem
for locations of buildings, \ac{POI}, land use information, and routing.
This approach significantly increases the transferability of the model.
Nonetheless, we still rely on household travel survey data for model calibration.
Once the model has been calibrated to a specific region, the model can be used by anyone without access to the original household travel survey.
Together with this paper, we publish a calibration that applies to Germany.
We aim to steadily increase its scope to more and more countries to ultimately achieve the goal of a
broadly applicable model that requires only \ac{OSM} data to run.

\section{Architecture}
\label{sec:Architecture}

\begin{figure}
    \centering
    \includegraphics[width=\textwidth]{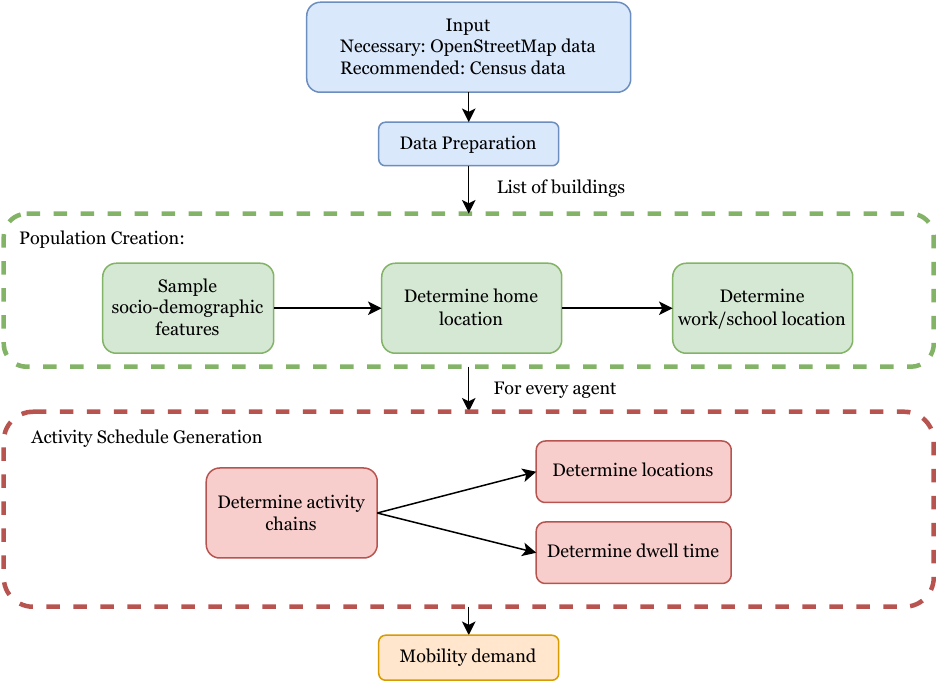}
    \caption{Architecture of \ac{OMOD}}
    \label{fig:Architecture}
\end{figure}

The following section will describe \ac{OMOD}'s architecture and the methodology that lead to the default parameterization.
The parameterization process of each model step is explained directly after its description.

The mobility demand generation process consists of three steps.
The first step is creating a model of the user-specified area (see \Cref{sec:DataPrep}).
This step involves parsing the \ac{OSM} file into a list of locations where activities can be conducted.
The second step is the creation of the population (see \Cref{sec:PopCreation}).
Here, the agents are assigned socio-demographic features,
and their inflexible locations (home, school, workplace) are determined. 
The third and last step handles the activity schedule generation (see \Cref{sec:MobDemGen}).
Here, the model determines what activities, where, and for how long every agent conducts on a given day.
This is the most complex step and takes up most of \ac{OMOD}'s runtime.

\Cref{fig:Architecture} depicts a high-level overview of \ac{OMOD}'s architecture.

\subsection{Data Preparation}
\label{sec:DataPrep}
The data preparation process parses the \ac{OSM} file and combines it with optional census data
to create a list of building instances
characterized by the features depicted in \Cref{tab:Building}.

\begin{table}
    \centering
    \begin{tabular}{ll}
        \toprule
        Features:\\
        \midrule
            coordinates             \\
            area                    \\
            population              \\
            landuse                 \\
            number of shops         \\
            number of offices       \\
            number of schools       \\
            number of universities  \\
            In focus area?          \\
        \bottomrule
    \end{tabular}
    \caption{Building features determined in the data preparation step.
    These features are parsed from the \ac{OSM} input data,
    except for \textit{population}, which is parsed from census data.}
    \label{tab:Building}
\end{table}

\begin{figure}
    \centering
    \includegraphics{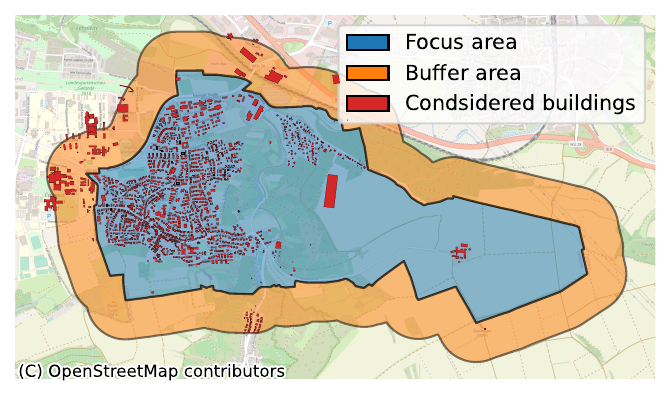}
    \caption{Example focus and buffer area in \ac{OMOD}. Depicted is Gerbrunn in Germany (\ac{OSM} id: 163738),
    with a buffer distance of \SI{500}{\meter}.}
    \label{fig:Area}
\end{figure}

Firstly, \ac{OMOD} determines the geometry of the area of interest (from here on called focus area) from a user-specified GeoJSON
file\footnote{Such a file can be easily created with tools like https://geojson.io.}.
Since people living in the immediate surrounding often significantly impact the mobility demand in the area of interest,
it is good practice to model the surrounding as well \cite{ortuzarModellingTransport4th2011}.
For this purpose, \ac{OMOD} implements the option to buffer the focus area by a given distance.
From here on, the additional area created this way will be called buffer area.
\Cref{fig:Area} shows an example of the area definition process.

Once the area is defined,
\ac{OMOD} parses all \ac{OSM} objects that intersect that area,
utilizing the \textit{Osmosis}\footnote{https://github.com/openstreetmap/osmosis} tool.
Objects with the \textit{building} tag are added to the building list.
The coordinates and area of each building are directly computed from the \ac{OSM} objects.
The remaining features are determined by combining the geometry information of each building with other \ac{OSM} information in the following manner:

To determine the land use feature,
we check whether each building intersects with a land use zone in the \ac{OSM} data.
The land use of the building is then that of the intersecting land use zone or \textit{none},
if none intersects.
Four land use classes are considered: \textit{residential}, \textit{industrial}, \textit{commercial}, and \textit{none}.
\textit{Residential} and \textit{industrial} are equivalent to the respective \ac{OSM} land use values.
\textit{Commercial} combines the \ac{OSM} land use values \textit{commercial} and \textit{retail}.
\textit{None} represents all other possible values in \ac{OSM}
and no specified land use.

We determine the features \textit{number-of-shops} and \textit{number-of-offices}
by counting the intersecting \ac{OSM} objects with the tag \textit{shop} or \textit{office}.
Similarly, the attributes \textit{number-of-schools} and \textit{number-of-universities} are determined by counting the intersecting \ac{OSM} objects
where the \textit{amenity} tag has the value \textit{school} or \textit{university},
respectively.

The population of each building is extracted from the optional census data.
This data must be formatted as a GeoJSON file containing a list of geometries and their populations.
For example, in Germany the Zensus 2011 \cite{statistischeamterdesbundesundderlanderZensus20112011} can be used,
where the population is given on the level of \SI{100}{\square\meter} cells.
The population of each census geometry is distributed uniformly across all intersecting buildings.
The population is assumed to be zero for buildings with no census data. 
If no census data file is provided,
\ac{OMOD} will not make assumptions about the number of inhabitants in each building.

\subsection{Population Creation}
\label{sec:PopCreation}

The population creation step creates a user-specified number of agents and defines their invariable attributes.
These include each agent's home location,
workplace,
and socio-demographic features.

\paragraph{Socio-demographic features}

First, the socio-demographic features are determined.
For every agent, a set of categorical features is sampled from a user-provided distribution.
If none is provided,
the model defaults to setting the features to \textit{undefined}.
This can be understood as defining the distribution to be the same as in the calibration survey.

It is assumed that one distribution of socio-demographic features is valid for the entire modeled area
(i.e., the socio-demographic makeup of the population does not differ significantly from district to district.).
The socio-demographic features considered by \ac{OMOD} are:

\begin{itemize}
    \item \textit{Age}. Possible values: \{0-40, 40-60, 60-100, or undefined\}
    \item \textit{Homogenous group}. Possible values: \{working adult, non-working adult, student, or undefined\}
    \item \textit{Mobility group}. Possible values: \{full car user, mixed car user, no car user, or undefined\}
\end{itemize}

These socio-demographic categories are chosen based on the analysis of Schlund \cite{schlundElectricVehicleCharging2021} and Joubert et al. \cite{joubertActivitybasedTravelDemand2020},
who determined that these features have the largest explanatory value for the mobility demand patterns observed in Germany and South Africa.
We use this somewhat limited number of categories because our approach for dwell time estimation (further explained in \Cref{sec:DwellTime})
needs a sufficient number of samples for every combination of socio-demographic features.
With an increasing number of features,
combinations increases exponentially,
and the maximum number of features that can be included is quickly reached.

\paragraph{Home location}

Each agent's home location is sampled from the list of buildings obtained in the data preparation step ($Buildings$).
The probability that building ($i$) is the home location of the agent
is the population of the building  ($POP_{i}$) divided by the total population in the modeled area:

\begin{equation}
	P(i=HOME) = \frac{POP_{i}}{\sum\limits_{\forall j \in Buildings}{POP_{j}}}
	\label{eq:HOME}
\end{equation}

If no census data is provided,
the home location is determined using the destination choice model of \ac{OMOD} (see \Cref{sec:DestChoice}).

\paragraph{Work/School location}

The work and school location sampling process depends on the realization of the home location.
With a given home location,
we determine the workplace/school location with a \ac{MNL} \cite{mcfadden1973conditional} that will also be used to choose flexible destinations (such as shopping locations).
The exact methodology will be explained in \Cref{sec:DestChoice}.
Broadly speaking, the model follows a disaggregated gravity model approach as described in \cite{ortuzarModellingTransport4th2011}.
The model comprises an attraction value estimated from building properties (see \Cref{tab:Building}) and 
a deterrence function based on the distance between the workplace/school and the home location.

\subsection{Activity Schedule Generation}
\label{sec:MobDemGen}

\begin{figure}
    \begin{CenteredBox}
\begin{lstlisting}[columns=fullflexible,language=json]
"activities": [
    {
        "type": "HOME",
        "stayTime": 327.073,
        "lat": 53.6157,
        "lon": 10.1072,
        "inFocusArea": true
    },
    {
        "type": "WORK",
        "stayTime": 591.966,
        "lat": 53.5256,
        "lon": 9.8951,
        "inFocusArea": true
    },
    {
        "type": "HOME",
        "stayTime": null,
        "lat": 53.6157,
        "lon": 10.1072,
        "inFocusArea": true
    }
]
\end{lstlisting}
\end{CenteredBox}
\caption{Example of an activity schedule produced by the activity schedule generation step.
\textit{type} states the activity category.
\textit{stayTime} describes how long (in minutes) the agent stays at the activity.
\textit{lat} and \textit{lon} indicate the location.
\textit{inFocusArea} states whether the activity was conducted inside the focus or buffer area (see \Cref{fig:Area}).
}
\label{fig:ActivitySchedule}
\end{figure}

The mobility demand generation step produces daily activity schedules for every agent in the population.
These schedules specify the number of activities the agent conducts on the day in question,
as well as their category, location, and duration (dwell time).
An example of such a schedule is depicted in \Cref{fig:ActivitySchedule}.

\ac{OMOD} first samples a chain of activities.
Then, the dwell times and locations are determined
conditionally on the outcome but independently of each other.

\ac{OMOD} uses a data-based approach to determine the activity chain and dwell times.
This simplifies the modeling process compared to traditional \ac{MNL} models
but limits number of socio-demographic features that can be included.
The destination choice model is implemented as a disaggregated gravity model (framed as an \ac{MNL}).
Therefore, it is easily expandable with additional explanatory features.

The activity schedule generation is calibrated with the German household travel survey \ac{MiD} \cite{infasMobilitatDeutschland2017}.
The \ac{MiD} is a large-scale reoccurring household travel survey published by the Federal Ministry of Digital and Transport.
We utilize the survey conducted in the year 2017.
This survey is distributed in the form of four data sets that differ based on spatial resolution and the level of detail of personal information.
For privacy preservation, the dataset with the highest spatial resolution contains the least detailed personal information and vice versa.
Of these datasets, we utilize the data set B3 with the highest spatial resolution but the least detailed personal information.
This data set contains information from about 300,000 respondents from 150,000 households.
For each respondent, the data set includes socio-demographic features,
access to different mobility options,
place of residence,
and a trip diary for one day in 2017.
In total, the dataset includes 1,000,000 trips.
For 500,000 of these, the start and stop locations are known with a resolution of $5km^2$,
and for 100,000 with the highest resolution of $500m^2$.

\subsubsection{Activity Chain}
\label{sec:ActivityChain}

The first step of the activity schedule generation process is sampling each agent's daily activity chain.
An activity chain describes the sequence of activities an agent undertakes on a given day.
For example, the chain $(home, \allowbreak work, \allowbreak shopping, \allowbreak home)$ states that the agent started his day at home,
went to work, then went shopping, and, finally, returned home.
Possible activity categories are \textit{home}, \textit{work}, \textit{school}, \textit{shopping}, and \textit{other}.

\begin{figure}
    \centering
    \includegraphics{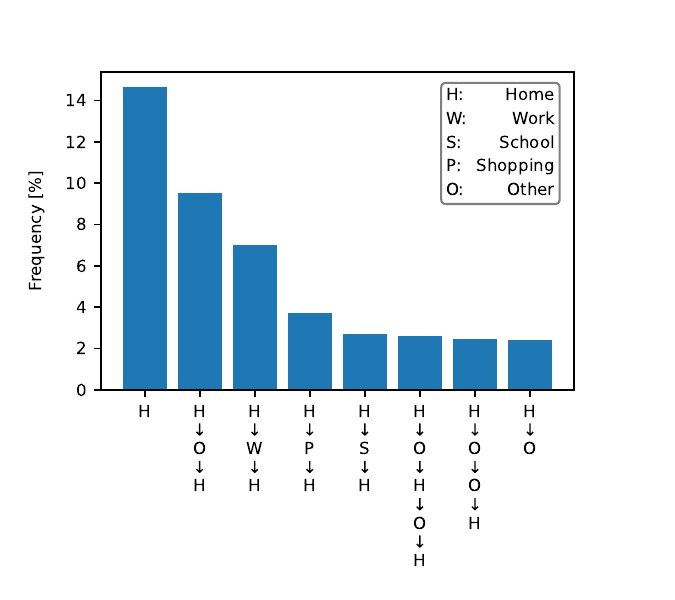}
    \caption{Empirical distribution of daily activity chains for an agent with \textit{undefined} socio-demographic features and for an \textit{undefined} weekday.
    For visual clarity, only the eight most common chains are depicted.}
    \label{fig:ActivityChain}
\end{figure}

We sample these activity chains directly from empirical distributions.
These are obtained by first filtering the \ac{MiD} for each socio-demographic feature and weekday combination
and then calculating the probability of each activity chain based on its frequency in the filtered dataset.
An example of such a distribution is depicted in \Cref{fig:ActivityChain}
for the case where all socio-demographic features and the weekday are \textit{undefined}.

For certain combinations of socio-demographic features,
only a few samples exist.
For example, students above the age of 60 are uncommon.
Therefore, we must ensure that the empirical distributions are based on adequate sample sizes.
We handle this issue by introducing a threshold for the minimum number of samples in the distribution.
The threshold is set to 30, based on the common rule of thumb \cite{hoggProbabilityStatisticalInference2006}.
If a distribution has a smaller sample size than this threshold,
we set individual socio-demographic features, or the weekday, to \textit{undefined} until an adequate sample size is reached.
This is done in the following order: 

\begin{gather*}
	Age \rightarrow Mobility \ group \rightarrow Homogenous \ group \rightarrow weekday
\end{gather*}

I.e., first, the age is set to \textit{undefined};
then, if the new distribution based upon this less restrictive set of conditions also has too few samples,
the mobility group is set to \textit{undefined},
and so on.
Since the entirely unrestricted distribution has enough samples,
this process will always return a distribution with adequate sample size.

The same threshold is applied to the minimum size of each activity chain.
This is necessary because later on, for each activity chain,
a distribution of dwell times is created (see \Cref{sec:DwellTime})
that again needs an adequate sample size.
If an activity chain does not have the necessary sample size,
it is removed from the empirical distribution.
Approximately 10\% of the samples must be discarded through this process.
After the removal, \ac{OMOD} includes 560 unique activity chains.
The longest remaining chains consist of up to 14 consecutive activities.

Since longer activity chains are more complex,
they are less likely to have enough samples,
causing an underestimation of the number of daily trips.
To combat this problem the probabilities of each activity chain are calibrated so
that the total probability of all chains with a given length is equal to the probability
of that length-group in the original dataset.

For consecutive days,
an additional condition on the distribution is that each day must start with the same activity the previous day ended with.
The first activity of the next day represents a continuation of the day's last activity.

\subsubsection{Dwell Time}
\label{sec:DwellTime}

With the activity chain for each agent determined, the time they spend on them can be sampled.

Similar to the process of activity chain sampling,
the dwell times are sampled from distributions fitted to the \ac{MiD}'s subset where the corresponding
socio-demographic features and weekday are present.
In this case, however,
the subset depends not only on the combination of socio-demographic features and weekday
but also on the specific activity chain.
In other words, one distribution exists for each socio-demographic feature, weekday, and activity chain combination.

Previous work often chose methodologies where the dwell times of activities are sampled conditionally
only on the dwell times at prior activities and not subsequent ones
\cite{schlundElectricVehicleCharging2021,gaete-moralesOpenToolCreating2021a}.
This underestimates how holistically individuals plan their entire day.
To combat this issue,
\ac{OMOD} samples dwell times from a multidimensional distribution,
where each dimension encodes an activity of the activity chain.
This way, all dwell times are sampled conjointly,
and the temporal information of each daily activity schedule is coherent.
This multidimensional distribution is modeled as a Gaussian Mixture.

For each feature combination we filter the \ac{MiD}'s accordingly and
fit a Gaussian Mixture to the filtered dataset using the scikit-learn python library.
The number of Gaussians/components in the mixture is determined by increasing the number of components
until the Bayesian Information Criterion score stops decreasing.

The \ac{MiD} does not specify when the last activity of a day ends.
We address this missing information by assuming that it lasts until midnight.
Therefore, the dwell time at the last activity is wholly determined by those of the other activities,
reducing the dimensionality of each mixture by one.
In the rare case that the last activity begins after midnight, its duration is zero.

\begin{figure}
    \centering
    \begin{subfigure}{\textwidth}
        \centering
        \includegraphics[scale=0.7]{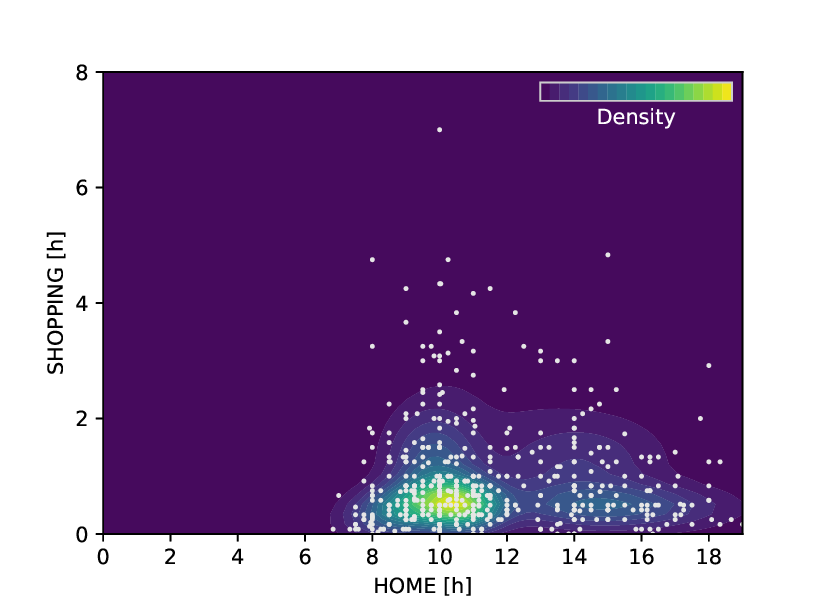}
        \caption{Activity chain: \textit{(home → shopping → home)}}
        \label{fig:GaussianMixture2D}
    \end{subfigure}
    \begin{subfigure}{\textwidth}
        \centering
        \includegraphics[scale=0.7]{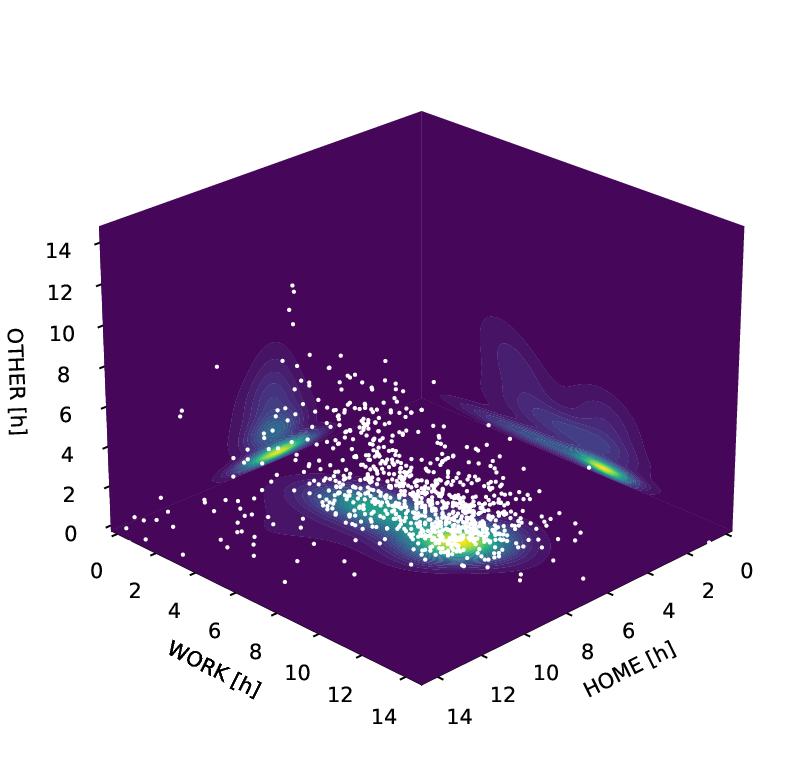}
        \caption{Activity chain: \textit{(home → work → other → home)}}
        \label{fig:GaussianMixture3D}
    \end{subfigure}
    \caption{Fitted Gaussian Mixture representing the dwell time distribution of two example activity chains with \textit{undefined} socio-demographic features
        on an \textit{undefined} weekday.
        The contours represent the density of the mixture projected onto each axis,
        where brighter colors represent a higher likelihood.
        The white points represent the records of the household travel survey (\ac{MiD}).
        For visual clarity, only a fraction of the records are depicted.
    }
    \label{fig:GaussianMixture}
\end{figure}

\Cref{fig:GaussianMixture} shows examples of Gaussian Mixtures resulting from the described process.
For example, the Gaussian depicted in \Cref{fig:GaussianMixture2D} describes the probability distribution for dwell times in the H→S→H chain.
A likely sample drawn from this distribution would be [10.5, 1.3],
meaning that the agent in question stays at home for 10.5 hours (i.e. until 10:30 AM),
then goes shopping for 1.3 hours,
and finally stayes at home again for the remainder of the day.

\subsubsection{Destination Choice}
\label{sec:DestChoice}

The destination choice step determines where agents conduct activities.
\ac{OMOD}s method for destination choice is based on the gravity model concept described by Ortuzar and Willumsen \cite{ortuzarModellingTransport4th2011}\footnote{Section 8.3.3}.
However, there are two key differences.

\begin{figure}
    \centering
    \includegraphics[width=\textwidth]{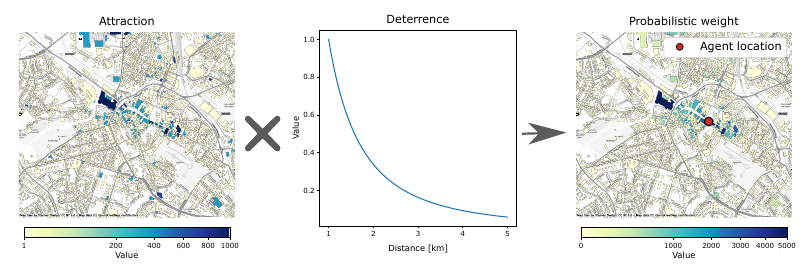}
    \caption{Destination choice: Architecture of the disaggregated gravity model.}
    \label{fig:DestChoice}
\end{figure}

Firstly, instead of aggregated traffic assignment zones,
the set of possible destinations comprises all buildings in the focus and buffer area.
Each time an agent has to choose a destination,
the entire set is considered.
Secondly, \ac{OMOD} can not rely on aggregated origin-destination information like \cite{ortuzarModellingTransport4th2011}
because the user is not required to provide survey data.
Therefore, we have to substitute this information.
This is done with the \textit{attraction} value $A_i$.
The \textit{attraction} governs the probability that a building is chosen when the agent's
location is taken into account.
It can be seen as the suitability of a building for a given purpose.
\ac{OMOD} determines this value based on land use and \ac{POI} information.
For example,
a building is more suitable for shopping if it contains shops.
Consequently, a building with shops has a higher \textit{attraction} value for shopping trips than one without.
The distance to the building is factored into the decision process through a deterrence function $f(d)$ in the same manner as in \cite{ortuzarModellingTransport4th2011}.

All taken together, the following equation describes the probability that building $i$ is chosen as the destination:

\begin{equation}
	P(i) = \frac{A_i \cdot f(d_{x, i})}{\sum\limits_{\forall j \in Buildings}{A_j \cdot f(d_{x, j})}}
	\label{eq:Gravity}
\end{equation}

or equivalently framed as an \ac{MNL}:

\begin{equation}
	P(i) = \frac{e^{V_{x,i}}}{\sum\limits_{\forall j \in Buildings}{e^{V_{x,j}}}}
	\label{eq:GravityMNL}
\end{equation}

with

\begin{equation}
	V_{x,i} = \ln(A_i) + \ln(f(d_{x, i}))
	\label{eq:GravityUtility}
\end{equation}

The distance ($d_{x, i}$) is calculated in reference to $x$,
which is either the home location
when the fixed locations (workplace and school) are determined
or the agent's current location.
This distance refers to the routed distance by car\footnote{
    In this regard, \ac{OMOD} neglects the aspect of mode choice.
    Some individuals prefer modes of transport other than the car,
    and some destinations are more easily reached with public transportation.
    In these cases, \ac{OMOD}'s deterrence associated with a particular location
    is falsely represented in \ac{OMOD}, leading to a misrepresentation of their probability.
    This problem is somewhat offset by the popularity of cars and the fact that currently,
    most destinations are most quickly reached by car \cite{infasMobilitatDeutschland2017}.
    Nonetheless, these errors will be significant for studies that want to evaluate populations that are less car-dependent.
    In future versions, we plan to incorporate this aspect of mode choice.
},
calculated with the open-source router GraphHopper\footnote{https://github.com/graphhopper/graphhopper}.

If the user provides no census data,
the home location is also determined with this model.
In that case, the value of the deterrence function is set to 1,
and only the attraction value of each building is relevant.

The parameterization of the deterrence function
and each building's \textit{attraction} depends on the activity conducted at the destination.
We call this activity the purpose of the trip from here on.
For example, a building has a different probability of being the agent's workplace than being the destination of a shopping trip.
The weekday and the socio-demographic group do not influence the deterrence function and \textit{attraction} value.

We obtain the parameterization of deterrence function and \textit{attraction} values with the \ac{MiD},
utilizing the methodology described in the following paragraphs.

\paragraph{Attraction}
\label{sec:attraction}

The \textit{attraction} of each building is estimated with the linear function depicted in \Cref{eq:AttractionFunc}
as inputs serve several \ac{OSM} features.
They can be separated into two groups.
The first group is comprised of the variables denoted by an $a$.
This group combines the area and land use of a building.
Depending on the land use, one of these variables equals the area of the building,
while the others are zero.
For example, if the building is in a residential area, $a_{Residential}$ equals the building's area
and $a_{Industrial}$, $a_{Commercial}$, and $a_{Other}$ are zero.
The second group comprises the variables denoted by an $u$ and describes the number of \ac{POI} associated with a building.
These can be shops, offices, schools, and universities.
Additionally, depending on the land use, either $u_{Residential}$,  $u_{Industrial}$, $u_{Commercial}$, or $u_{Other}$ equals one and the others zero.

\begin{equation}
    \begin{aligned}
        A_{i} = 
        & \ 1 + \theta_0 \cdot a_{Residential} + \theta_1 \cdot a_{Industrial} + \theta_2 \cdot a_{Commercial} + \\
        & \ \theta_3 \cdot a_{Other} + \theta_4 \cdot u_{Office} + \theta_5 \cdot u_{Shops} + \\
        & \ \theta_6 \cdot u_{Schools} + \theta_7 \cdot u_{Universities} + \theta_8 \cdot u_{Residential} + \\
        & \ \theta_{9} \cdot u_{Industrial} + \theta_{10} \cdot u_{Commercial} + \theta_{11} \cdot u_{Other}
    \end{aligned}
	\label{eq:AttractionFunc}
\end{equation}

We determine the coefficients $\theta_k$ ($k \in \{0,1, ..., 11\}$) of \Cref{eq:AttractionFunc} through maximum likelihood estimation.
The \textit{attraction} value $A_i$ describes the probabilistic weight that a building $i$ is the destination
for a trip under the condition that the agent's location is unknown.
Therefore, the probability that a trip with an unknown origin ends at building $i$ is:

\begin{equation}
    P(i) = \frac{A_i}{\sum\limits_{\forall j \in Buildings}{A_j}}
	\label{eq:FitAttraction}
\end{equation}

The set \textit{Buildings} describes all possible locations that an agent could have chosen as a destination.
Since the \ac{MiD} is a Germany-wide study, \textit{Buildings} is comprised of all buildings in Germany.
However, the \ac{MiD} only specifies trip destinations at a resolution of \SI{500}{\square\meter}.
Therefore, we aggregate all the features in \Cref{eq:AttractionFunc} on the level of the \SI{500}{\square\meter} cells
and consider all of these cells as suitable destination choices.
This aggregation is possible due to the linearity of the \textit{attraction} equation.

With the choice set defined,
we can find the parameters $\theta_k$ of \Cref{eq:AttractionFunc}
that maximize the probability of the observed trip destinations in the \ac{MiD} for the trip purpose. 
The \textit{L-BFGS-B} solver implementation of the python package \textit{SciPy} is used to determine the maximum of the likelihood function\footnote{
https://docs.scipy.org/doc/scipy/reference/optimize.minimize-lbfgsb.html}.

We set the lower bounds of $\theta_k$ to zero.
This means that each \ac{OSM} feature can only attract but never deter.
The introduction of these bounds has two main benefits.
Firstly, they ensure that \Cref{eq:FitAttraction} always yields positive non-zero probabilities for every building.
Secondly, they introduce a level of regularization to the model,
reducing the number of variables.

We further reduce the number of features with the following methodology.
First, we fit the model several times, each time using all features except one,
meaning that the coefficient $\theta$ of one feature is fixed to zero.
Then we compare the likelihood of these iterations and create a ranking of features
based on how much their absence worsened the results.
After that, we build a model with only the most important feature,
then one with only the two most important features,
and so on, until no significant increase in likelihood is observable.
The last set of features that improved the model is then chosen.

\begin{table}
    \centering
    \begin{tabular}{lrrrrrr}
        \toprule
        Activity & $\theta_0$ & $\theta_4$ & $\theta_5$ & $\theta_6$ & $\theta_7$\\
        \midrule
            \textit{home}     & 0.0327 &       0 &  314.09 & 1679.18 &       0 \\
            \textit{work}     &      0 &  727.14 &  280.69 &  611.39 &       0 \\
            \textit{school}   &      0 &  339.04 &  132.36 & 2115.64 & 3061.74 \\
            \textit{shopping} &      0 &       0 &  348.44 &       0 &       0 \\
            \textit{other}    & 0.0370 & 2789.23 & 2179.04 & 1966.55 &       0 \\
        \bottomrule
    \end{tabular}
    \caption{Coefficients of the \textit{attraction} function.
    See \Cref{eq:AttractionFunc} for the \ac{OSM} feature corresponding to each $\theta_k$.
    $\theta$s that are not shown are zero.}
    \label{tab:attractionTheta}
\end{table}

\Cref{tab:attractionTheta} shows the resulting $\theta_k$ for each trip purpose.
Since a building without any features has a fixed probability weight of one,
the results can be interpreted as how much more likely a building with a specific \ac{POI} is compared to a generic building.
For example, a building with a shop is 350 times more likely to be a shopping destination.
Similarly, we can interpret the coefficients of the residential area variable ($\theta_0$),
which states how the probability increases with the area of the building (unit: square meter).

The results indicate that primarily \ac{POI} data increases the \textit{attraction} of a building,
suggesting that these features have higher explanatory value for predicting trip destinations.
However, during the model creation phase,
we tested several different architectures for the \textit{attraction} function.
Land use and area features were more important for many similarly well-performing iterations.
Therefore, we can not definitively say which features are the most relevant.
Presumably, the high correlations between features on the aggregation level of \SI{500}{\square\meter}
are responsible for the inconclusiveness of our results.

Some artifacts of the aggregation can still be seen in the results for the \textit{home} activity,
where shops and schools have a significant impact.
Note that the model for the \textit{home} purpose is only chosen in the absence of census data.
These artifacts are the main reason for introducing the feature reduction processes described above.
Overall, however, the coefficients take reasonable values.
The \textit{school} purpose in the \ac{MiD} includes apprenticeships.
Therefore, the non-zero coefficient of shops and offices is not unreasonable.

\paragraph{Deterrence function}
\label{sec:deterrence}

For the parameterization of the deterrence function we evaluate the following functional forms for $f(d_{x, j})$:

\begin{center}
    \begin{tabular}{lll}
        exponential (E) & $f(d_{x, j}) = exp(\beta d_{x,j})$ \\
        power \& exp. (PE) & $f(d_{x, j}) = d_{x,j}^{n} \cdot exp(\beta d_{x,j})$ \\
        lognormal (L) & $f(d_{x, j}) = \frac{1}{d_{x,j}\sigma\sqrt{2\pi}} exp(-\frac{(ln(d_{x,j})-\mu)^2}{2\sigma^2})$ \\
        lognormal \& exp. (LE) & $f(d_{x, j}) = \frac{1}{d_{x,j}\sigma\sqrt{2\pi}} exp(-\frac{(ln(d_{x,j})-\mu)^2}{2\sigma^2})\cdot exp(\beta d_{x,j})$  
    \end{tabular}
\end{center}

The first two functional forms are commonly used in related work \cite{ortuzarModellingTransport4th2011}.
The lognormal form is evaluated
because it closely resembles the trip distance distribution in the \ac{MiD}.
Finally, the combined lognormal and exponential distribution is an attempt to introduce better tail behavior to the lognormal functional form.

In \Cref{eq:GravityMNL}, the deterrence function always occurs inside a logarithm.
Therefore, we can significantly simplify the parameterization step of the deterrence function by directly fitting the logarithm.

If we take the natural logarithm of the functional forms, we get:

\begin{center}
    \begin{tabular}{ll}
        E  & $ln(f(d_{x, j})) = \beta d_{x,j}$ \\
        PE & $ln(f(d_{x, j})) = \beta d_{x,j} + nln(d_{x,j})$ \\
        L  & $ln(f(d_{x, j})) = -\frac{1}{2\sigma^2}ln^2(d_{x,j}) + (\frac{\mu}{\sigma^2} - 1)ln(d_{x,j}) + \frac{\mu^2}{2\sigma^2} - ln(\sigma\sqrt{2\pi})$ \\
        LE & $ln(f(d_{x, j})) = -\frac{1}{2\sigma^2}ln^2(d_{x,j}) + (\frac{\mu}{\sigma^2} - 1)ln(d_{x,j}) + \frac{\mu^2}{2\sigma^2} - ln(\sigma\sqrt{2\pi}) + \beta d_{x,j}$
    \end{tabular}
\end{center}

If we disregard constant terms and aggregate the coefficients of each linear term to individual independent parameters,
we get the following linear forms:

\begin{center}
    \begin{tabular}{ll}
        E  & $ln(f(d_{x, j})) = \vartheta_0 d_{x,j}$\\
        PE & $ln(f(d_{x, j})) = \vartheta_0 d_{x,j} + \vartheta_1ln(d_{x,j})$ \\
        L  & $ln(f(d_{x, j})) = \vartheta_0ln^2(d_{x,j}) + \vartheta_1ln(d_{x,j})$ \\
        LE & $ln(f(d_{x, j})) = \vartheta_0ln^2(d_{x,j}) + \vartheta_1ln(d_{x,j}) + \vartheta_2d_{x,j}$
    \end{tabular}
\end{center}

These are less constraint versions of the original functional forms that are significantly easier to handle in the parameterization step.

For each functional form,
we find the parameters that maximize the likelihood of trip destinations in the \ac{MiD},
assuming that each destination's probability is governed by \Cref{eq:GravityMNL}.
The parameters of the \textit{attraction} function are already determined in the previous step and assumed constant here.

Similar to the parameterization of the \textit{attraction} function,
we consider all buildings in Germany as possible destinations and have to aggregate their \textit{attraction} to \SI{500}{\square\meter} cells.
Additionally, we need the distance between the trip's origin and all possible destinations.
We utilize GraphHopper to determine the routed distance (by car) between every cell centroid
and every other cell centroid.

This routing computation is very time intensive;
even then, we leverage the ShortestPathTree API of GraphHopper.
Therefore, the distance matrix has to be precomputed and stored in memory to reach reasonable optimization times.
However, the memory consumption would be unacceptable with a distance matrix of the size $(1.5 \cdot 10^6)^2$ (the number of \SI{500}{\square\meter} cells in Germany squared).
For this reason, three simplifications are necessary.
Firstly, we only determine the distance between cells that enclose buildings, halving the number of cells.
Secondly, we introduce a distance limit of \SI{300}{\kilo\meter}.
If the routed distance between two cells is above this limit,
we do not route.
Instead, we substitute with a beeline calculation fast enough not to require precomputation.
Thirdly, we digitize the distance information into the \SI{50}{\meter} wide
bins.
The process of this digitization is described in more detail in \ref{sec:smplDestChoice}.

With these simplifications, all the necessary data needed for the maximum likelihood estimation can be stored in memory,
and the optimal parameters of each functional form are determined.
The functional form is chosen for each trip purpose where the trip distance distribution
produced by \Cref{eq:GravityMNL} is closest to that in the \ac{MiD}.
We use the Kolmogorov-Smirnov test to measure the goodness of fit.

\begin{table}
    \centering
    \begin{tabular}{lrrrr}
        \toprule
            Activity            & Form & $\vartheta_0$ & $\vartheta_1$ & $\vartheta_2$ \\
        \midrule
            \textit{work}       & EP & -0.035 & -0.919 & - \\
            \textit{school}     & LE & -0.235 & -1.176 & 0.005 \\
            \textit{shopping}   & L  & -0.215 & -1.414 & - \\
            \textit{other}      & L  & -0.180 & -1.067 & - \\
        \bottomrule
    \end{tabular}
    \caption{Coefficients and functional forms of the deterrence function for each trip purpose (activity at destination).
    The unit of distance used in the functions is kilometer.}
    \label{tab:DeterrenceTheta}
\end{table}

\begin{figure}
    \centering
    \includegraphics{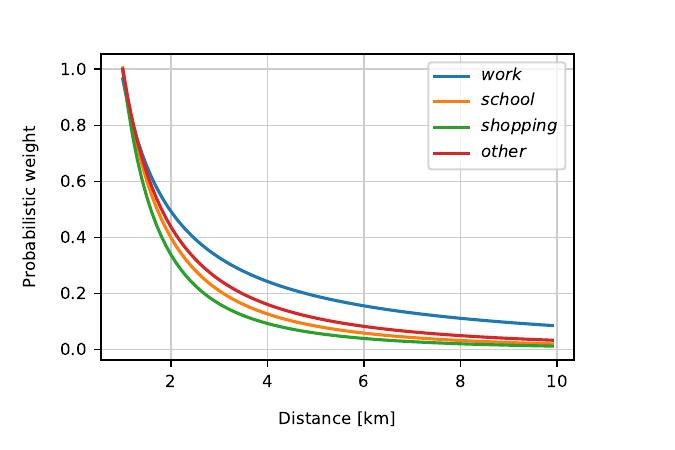}
    \caption{Values of the fitted deterrence functions $f(d_{x, j})$ over the distance from origin to destination.}
    \label{fig:DeterrenceFn}
\end{figure}

The process results in the deterrence function parameterization depicted in \Cref{tab:DeterrenceTheta}\footnote{The
deterrence function of \textit{school} has its minimum at \SI{827}{\kilo\meter} and subsequently increases again.
Since it makes no theoretical sense that probabilities start to rise again at very long distances,
the probability of choices beyond the minimum are set to zero.
The minimum can be explained by the absence of choices with higher distances in the training data.}. 
\Cref{fig:DeterrenceFn} shows how the probabilistic weight of a destination falls over distance.
We can see that the general shape of all deterrence functions is similar.
However, the rate of decline differs significantly between purposes.

\paragraph{Grid}
\label{sec:Grid}

During runtime,
\Cref{eq:GravityMNL} has to be evaluated for every building in the model area
each time an agent conducts an activity with no fixed location.
This involves calculating the distance to every building and constitutes \ac{OMOD}'s main performance bottleneck.

To speed up this process, we introduce a grid.
When an agent makes a destination choice,
first, the probability of each grid cell is determined
using \Cref{eq:GravityMNL}, but with the aggregated \textit{attraction} of all buildings within the grid cell
and the distance from the agent to the centroid of these buildings.
Subsequently, a cell is sampled,
and the building inside it is chosen solely based on its attraction value,
disregarding the distance differences of buildings within the same cell.

This grid can be defined in various ways.
The naive approach is a regular grid where all grid cells are squares of equal size.
However, since \ac{OMOD}s runtime rises quadratically with the number of cells
(for every trip with a flexible location we have to calculate the distance from a cell to every other cell),
it is advisable to introduce a more efficient grid.
The error the grid introduces is characterized by the average distance between a building and the centroid representing its associated grid cell.
This is the case because we calculate the distance from the origin only to the centroid of each cell,
neglecting the positional deviation between the cell centroid and building location.
Therefore, the best grid with $k$ cells is that where the sum of the within-cell variance of the building positions is smallest.
We can find a suitable grid with the k-Means algorithm.

\begin{figure}
    \centering
    \includegraphics{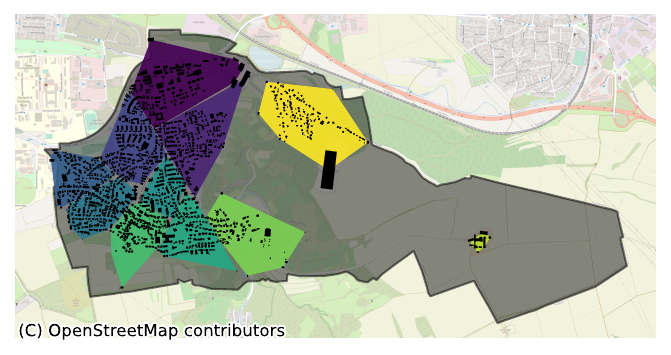}
    \caption{Example of the grid. Depicted is Gerbrunn, Germany (OSM id: 163738).}
    \label{fig:Grid}
\end{figure}

There are two problems with this approach.
Firstly, the runtime of the standard k-Means algorithm is not insignificant (several minutes in our trials).
Secondly, the number of cells should not be constant but increase with the size of the model area.
We solve both of these issues by using the Bisecting-K-Means variant of the algorithm.
This version results in a slightly higher within-cell variance but has significantly lower runtimes
and has the added advantage that it can be implemented with a custom stopping criterion.
Instead of stopping once we reach $k$ clusters/cells,
we terminate the clustering algorithm once
the average distance between each building and its associated centroid falls below a fixed threshold,
representing the resolution of the grid.
Per default, this threshold is \SI{150}{\meter}.
In the subsequent validation, we use the default resolution for all tests.
The resulting grid is depicted in \Cref{fig:Grid}.

The presented grid creation process greatly reduces the necessary number of grid cells compared to a regular grid.
Nonetheless, for large areas, the number of cells can quickly reach limits where the computation time becomes impractical.
This issue is especially problematic since large buffer areas are often necessary to accurately recreate individuals' daily driving distances.
We combat this issue by reducing the grid resolution with the distance from the focus area,
meaning that buildings far away from the original focus area are grouped into larger and larger cells.
Specifically, we run the clustering approach described above separately for groups of buildings.
The first group is all buildings in the focus area;
the second is all buildings in the buffer area with a distance of fewer than \SI{10}{\kilo\metre} to the focus area,
then those between \SI{10}{\kilo\metre} and \SI{20}{\kilo\metre},
and so on.
We are doubling the grid resolution threshold for every new group of buildings.
This way, we achieve adequate runtimes for large buffer radii.

\section{Validation}
\label{sec:validation}

We validate the model by determining how well the model reproduces
the observations of the German household travel survey \ac{MiD} \cite{infasMobilitatDeutschland2017}.
First, we determine how close the spatial patterns of the mobility demand are reproduced.
To do so,
we compare how many trips each zone attracts,
the origin-destination matrices,
and the daily driven distances.
Secondly, we analyze how well temporal patterns are reproduced
by evaluating the share of persons that conduct a specific activity over the course of a week.

Please note that this validation is somewhat limited by the fact
that the \ac{MiD} is our primary source for calibration data.
Therefore, the test and train sets are not strictly separated.
However, regarding destination choice, we reduced the entire information of the \ac{MiD}
to the parameters described in \Cref{tab:attractionTheta,tab:DeterrenceTheta},
in total, 24 non-zero parameters.
Consequently, the risk of overfitting is reduced.
The temporal characteristics have a significantly larger number of parameters
that are also less explainable.
However, the Gaussian Mixture methodology generally resulted in low numbers of Gaussians
and the risk of overfitting to outliers is small.
Regardless, the current parameterization of \ac{OMOD} can only reproduce mobility behavior as observed in the \ac{MiD}.

\subsection{Spatial Validation}
\label{sec:SpatialVal}

We evaluate the spatial model performance for three differently sized German cities.
Kassel a smaller city with 200,000 inhabitants,
the agglomeration area of Nuremberg with 1.3 million inhabitants,
and Hamburg a large city with 2.5 million inhabitants.
We chose these cities because of their populations and location (north, middle, and south).

Every city is simulated with 100,000 agents
for four \textit{undefined} days,
using data from the German census of 2011 \cite{statistischeamterdesbundesundderlanderZensus20112011}
to determine the distributions of \textit{home} locations.
The administrative boundary of each city defines the focus areas\footnote{
The north sea territory of Hamburg is ignored.
For the Nuremberg agglomeration area, the areas of the cities Nuremberg, Erlangen, and Fürth are combined.
}.
Each focus area is buffered with a distance of \SI{40}{\kilo\meter}.

\subsubsection{Zonal Trip Attraction}

\begin{figure}
    \begin{subfigure}{\textwidth}
        \centering
        \includegraphics{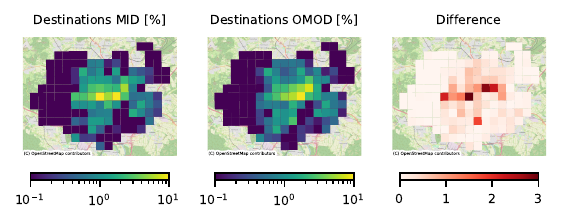}
        \caption{Kassel}
        \label{fig:AttractionKassel}
    \end{subfigure}
    \begin{subfigure}{\textwidth}
        \centering
        \includegraphics{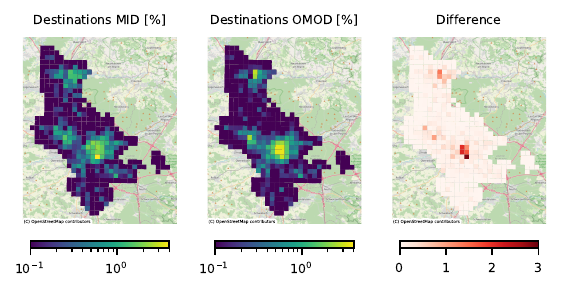}
        \caption{Nuremberg}
        \label{fig:AttractionNuremberg}
    \end{subfigure}
    \begin{subfigure}{\textwidth}
        \centering
        \includegraphics{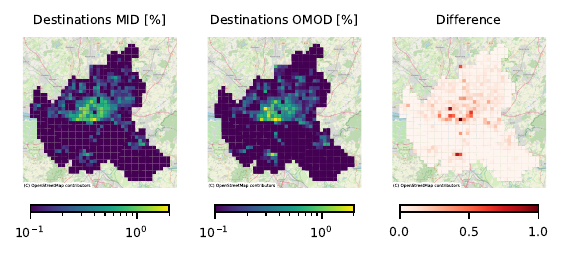}
        \caption{Hamburg}
        \label{fig:AttractionHamburg}
    \end{subfigure}
    \caption{Comparison of the share of trips that end in each \SI{1}{\kilo\meter} cell between \ac{OMOD} and the household travel survey \ac{MiD}.}
    \label{fig:Attraction}
\end{figure}

This test compares the trip destination distributions between the survey and model
based on the share of trips that end in a given zone,
where zones are cells of grids with \SI{500}{\meter}, \SI{1}{\kilo\meter}, and \SI{5}{\kilo\meter}
resolution (these grids are strictly for validation purposes and not to be confused with the grid used by \ac{OMOD} internally).
The results are depicted in \Cref{fig:Attraction} for each focus area and with a resolution of \SI{1}{\kilo\meter}.
In \ref{sec:appendixAttraction}, quantitive results are displayed for all resolutions.

Qualitative results show that OMOD reproduces the overall popularity of different city districts for all three city sizes.
This observation is supported by the high $R^2$ values of 0.87 to 0.95
on the lowest resolution level.
For the medium resolution (\SI{1}{\kilo\meter}),
the model's performance decreases to 0.61-0.77
and,
for the highest resolution (\SI{500}{\meter}),
to 0.22-0.36.
These reductions can be ascribed to the fact that predicting mobility demand
with higher spatial resolution becomes increasingly difficult.

For each trip purpose, the results show similar performance
and a similar performance decline for higher resolutions as in the overall case. 
The exception is \textit{home}, where the performance declines significantly slower,
reaching an $R^2$ value of 0.5 at the highest resolution.
This result is unsurprising,
as the validation uses census data with \SI{100}{\meter} resolution.
The remaining error is likely because of the six-year gap between the creation of the census and the household travel survey.
If we do not use census information, we get $R^2$ values of 0.81-0.93 for the lowest resolution,
with a similiar decline in performance at higher resolutions compared to the other purposes.

\subsubsection{Origin-Destination Matrix}

This test determines how well the origin\Hyphdash destination matrix of the focus area is reproduced.
We use the same methodology as in the trip attraction evaluation,
only here, the probability distributions describe the probability
that a trip starts in one cell and ends in another.
While the first test determined whether the relative popularity of different parts of the city is well represented,
this test determines whether the flows between city parts are realistic.
We evaluate the $R^2$, the mean absolute error (MAE), and the Jensen-Shannon divergence
between the flow distributions of the survey and model.
The \SI{500}{\meter} resolution level is not evaluated because,
with an average number of 50 buildings per zone and more than 500 possible destinations (even for the smallest city),
we enter the realm where we would need to predict the behavior of individuals,
something that activity based models are incapable of \cite{davidsonSynthesisFirstPractices2007}.
The results are depicted in \Cref{tab:OD}.

\begin{table}
    \centering
    \begin{tabular}{llrrr}
        \toprule
        City & Resolution & $R^2$ & MAE [\%] &  Jensen-Shannon \\
        \midrule
        Kassel    &  \SI{5}{\kilo\meter}  & 0.956 &                0.211 &           0.134 \\
                  &  \SI{1}{\kilo\meter}  & 0.461 &                0.005 &           0.506 \\
        Nuremberg &  \SI{5}{\kilo\meter}  & 0.715 &                0.044 &           0.174 \\
                  &  \SI{1}{\kilo\meter}  & 0.177 &                0.001 &           0.576 \\
        Hamburg   &  \SI{5}{\kilo\meter}  & 0.868 &                0.017 &           0.206 \\
                  &  \SI{1}{\kilo\meter}  & 0.025 &                0.000 &           0.626 \\
        \bottomrule
    \end{tabular}
    \caption{Origin-destination evaluation}
    \label{tab:OD}
\end{table}

For all cities the results on the \SI{5}{\kilo\meter} resolution are good.
This is promising as the origin-destination matrix has $n^2$ entrees, where $n$ is the number of grid cells.
Therefore, it is significantely more complex than the trip destination distribution.

The model underperforms for Nuremberg,
primarily because of a significant overestimation of trips that start and end in the city center.
We can trace this back to a very high density of \ac{POI} there.
Possibly, at a certain density,
the trip attraction increase of additional \ac{POI} diminishes,
suggesting that the \textit{attraction} function could benefit from the addition of saturation effects.

The results on the \SI{1}{\kilo\meter} resolution suffer from sample size issues.
On this resolution, the survey contains one record for every five origin-destination pairs in Kassel,
one for every twelve in Nuremberg,
and one for every 33 in Hamburg.
Nonetheless,
even with limited samples,
it seems clear that \ac{OMOD} could perform better in this regard.
Reproducing origin-destination matrices on this resolution is difficult,
espacially if the model is not fine-tuned to the region in question.
As a way forward,
we suspect increasing the number of explanatory features
by adding georeferenced census data in combination with more detailed socio-demographic features
can lead to a significant performance increase.
However,
the core use case of \ac{OMOD} should always utilize data available to anyone.
Therefore,
optimizing the model based on optional additional data sources has little priority.
Another likely source of error is, neglecting congestion and modes other than the car (in particular public transportation) in the destination choice step,
which is responsible for a significant misrepresentation of the generalized cost of travel for several origin-destination pairs.
Implementing these into \ac{OMOD} will be less problematic regarding ease of use
because the GTFS format provides a good de facto standard for public transport timetables.
However, an implementation poses significant runtime issues that need to be addressed.

\subsubsection{Daily Driven Distance}
\label{sec:dailyDrivenDistance}
Another crucial descriptive metric of mobility demand is the daily driven distance of individuals.

\begin{table}
    \centering
    \begin{tabular}{llrr}
        \toprule
        City & Metric &    MiD &   OMOD \\
        \midrule
        Kassel      & 0.25-quantile   &  3.837 &  2.724 \\
                    & Median          & 12.115 & 10.478 \\
                    & 0.75-quantile   & 25.650 & 21.797 \\
                    & Mean            & $21.869 \pm 1.74 $ & $18.298 \pm 0.06 $ \\
        Nuremberg   & 0.25-quantile   &  4.200 &  3.040 \\
                    & Median          & 13.320 & 12.200 \\
                    & 0.75-quantile   & 29.480 & 28.132 \\
                    & Mean            & $23.301 \pm 0.90 $ & $20.668 \pm 0.07 $ \\
        Hamburg     & 0.25-quantile   &  4.900 &  3.684 \\
                    & Median          & 14.400 & 16.228 \\
                    & 0.75-quantile   & 30.400 & 36.175 \\
                    & Mean            & $24.664 \pm 0.78 $ & $24.990 \pm 0.08 $ \\
        \bottomrule
    \end{tabular}
    \caption{Daily kilometer comparison with a buffer distance of \SI{40}{\kilo\meter}}
    \label{tab:DailyDistance}
\end{table}

\ac{OMOD} does not specify the route an agent takes from A to B.
For the validation, we assign routes with an all-or-nothing approach,
always choosing fastest route by car\footnote{As determined by GraphHopper} and disregarding congestion effects.
This simple assignment strategy certainly comes with its own error that can not easily be separated from the
error inherent to \ac{OMOD}.
The results are summarized in \Cref{tab:DailyDistance}.
Note
that the \ac{MiD}'s results do not include regular trips conducted during work (for example, if the person is a postman)
or trips conducted while on vacation
since \ac{OMOD} does not model these kinds of trips.

We can see that for the largest city,
the average daily kilometers are very closely reproduced.
However, a slight overestimation of the variance occurs,
meaning that short and long trips occur too often
but in such a way that the average is preserved.

\begin{figure}
    \centering
    \includegraphics{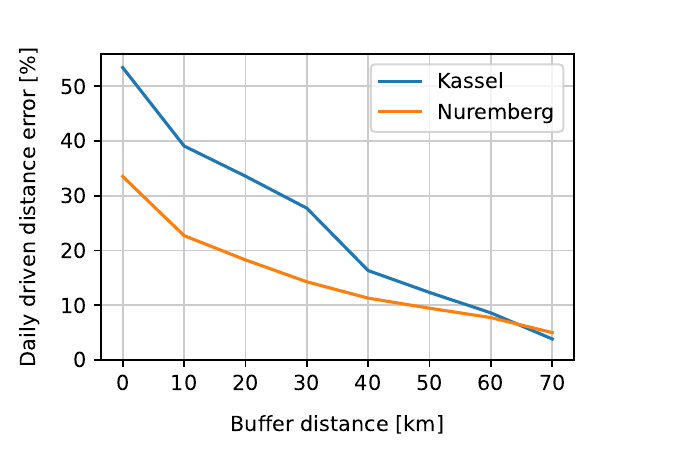}
    \caption{Dependency of the average daily driven distance error on the size of the buffer area.}
    \label{fig:BufferSensitivity}
\end{figure}

For the smaller cities, we get an underestimation of the average daily kilometers.
To some extent, this error is inevitable.
Since only buildings included in the model can be destinations,
the average daily distance can not exceed the average number of trips
multiplied by the furthest distance in the model area.
With an increased buffer area, the error should decrease.
That indeed happens,
as can be seen in \Cref{fig:BufferSensitivity}.
With an increased buffer distance,
the error for Nuremberg and Kassel falls to below \SI{5}{\percent}.
The error could be reduced further.
However, increasing the buffer area further is increasingly costly
due to the quadratic increase in routing calculations that have to be done (see \Cref{sec:Grid}).

\subsection{Temporal Validation}
\label{sec:TemporalVal}

We have already ascertained the spatial validity of \ac{OMOD}.
Additionally, the model should reproduce the temporal patterns of real mobility demand.
To validate this,
we compare the share of agents conducting a specific activity at each point in time over a week in \ac{OMOD} and the survey.

Some notes about our methodology:
Firstly,
since in \ac{OMOD}, the first day for all agents begins at home, we simulate two weeks,
the first to let the model settle and the second to use in the actual comparison.
Secondly,
trip assignment is conducted with the same all-or-nothing method as in \Cref{sec:dailyDrivenDistance}.

\begin{figure}
    \centering
    \includegraphics{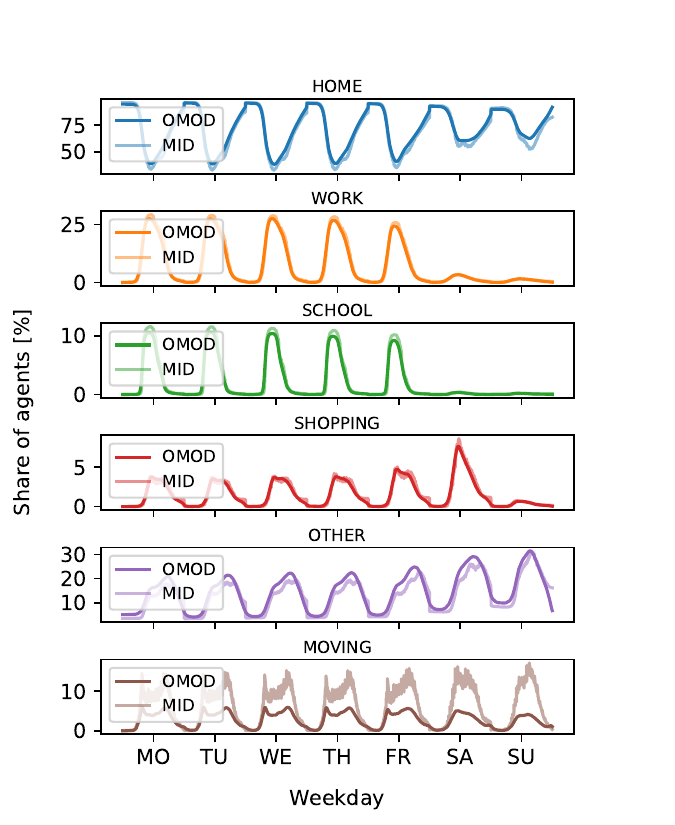}
    \caption{Temporal validation: Depicted is the share of agents conducting a specific activity over the course of a generic week
             in \ac{OMOD} and the \ac{MiD} household travel survey.}
    \label{fig:Temporal}
\end{figure}

\Cref{fig:Temporal} depicts the results.
Overall, the temporal aspects of the mobility demand are very well reproduced.
At no point did more than \SI{13}{\percent} of the agents conduct the wrong activity.
The average error over the timespan is \SI{5}{\percent}.
This error is caused almost entirely by underestimating the number of moving agents,
which can be traced back to the trip assignment process used in validation.
In the all-or-nothing assignment process,
all trips are conducted by car, 
only the pure driving duration is considered,
and congestion effects are disregarded.
On the other hand,
the \ac{MiD} includes all modes of transport, congestion effects, and inefficiencies like parking spot searches.
Because of these factors, the all-or-nothing approach underestimates travel times,
leading to an underestimation of moving agents and a corresponding overestimation at some other activity.
Additionally, since \ac{OMOD} does not specify fixed start times for activities the shorter travel times
also mean that activities start and end earlier than they should,
leading to a slight left shift of all activities that is corrected at the end of each day.

In the \ac{MiD} time series, discontinuities are visible at midnight.
These stem from the fact that each person is only questioned about one day
from waking up to midnight.
\ac{OMOD} smooths these discontinuities out because it must remain self-consistent,
leading to a discrepancy between the model and survey that should not be regarded as an error.

\section{Runtime}
\label{sec:Runtime}

\ac{OMOD}s runtime is mainly influenced by four components of the program.
These are: parsing the \ac{OSM} file, GraphHopper initialization,
the routing matrix's pre-computation,
and the main simulation (where the simulation steps described in \cref{sec:MobDemGen} are run for all agents).
In this section, we evaluate the runtime of each of these components for the focus area of Nuremberg,
defined with the same boundaries as in \Cref{sec:validation}.
All tests are run on an ordinary scientific laptop (CPU: i7-1165G7 @ 2.8 GHz, RAM: 16 GB).

\begin{table}
    \centering
    \begin{tabular}{ll|rr}
        \toprule
        \ac{OSM} file: &&& \\
        NUTS - Level & Name & \ac{OSM} parsing & GraphHopper init \\
        \midrule
        NUTS - 2 & Middle Franconia  & 1min 10s & 9s \\
        NUTS - 1 & Bavaria           & 1min 29s & 1min 2s \\
        NUTS - 0 & Germany           & 3min 30s & 6min 26s  \\
        \bottomrule
    \end{tabular}
    \caption{Runtimes of \ac{OSM} parsing and GraphHopper initialization for differently sized \ac{OSM} files.}
    \label{tab:RuntimeOfInit}
\end{table}

The runtime of the first two components (\ac{OSM} parsing and GraphHopper initialization)
is dependent on the size of the \ac{OSM} file.
\Cref{tab:RuntimeOfInit} depicts runtimes for \ac{OSM} files of three differently sized regions encompassing the example city.
The runtimes of these components are acceptable;
\ac{OSM} files of entire countries can be parsed in a reasonable time.
The results are stored and reused for all subsequent runs of the same area.

\begin{table}
    \centering
    \begin{tabular}{l|r|r|rrrr}
        \toprule
        Buffer dist. & Area size    & Routing  & \multicolumn{4}{c}{Main Simulation. \#Agents:} \\
        Unit: $km$      & Unit: $km^2$ & matrix init & $10^3$ & $10^4$ & $10^5$ & $10^6$     \\
        \midrule
        0  &  $3.3 \cdot 10^2$ &  2min 44s & 1s & 4s & 29s & 4min 56s \\
        10 &  $1.7 \cdot 10^3$ &  8min 42s & 1s & 4s & 36s & 6min  9s \\
        20 &  $3.7 \cdot 10^3$ & 17min 37s & 1s & 6s & 42s & 7min 16s \\
        30 &  $6.2 \cdot 10^3$ & 32min 50s & 1s & 6s & 47s & 7min 55s \\
        40 &  $9.3 \cdot 10^3$ & 52min 10s & 2s & 7s & 49s & 8min 27s \\
        \bottomrule
    \end{tabular}
    \caption{
        Runtimes of routing matrix precomputation and the main simulation.
        The main simulation is the only component of the runtime that can not be stored and reused on subsequent runs.
    }
    \label{tab:Runtime}
\end{table}

As stated in \Cref{sec:Grid}, the main performance bottleneck of \ac{OMOD} is repeatedly calculating the
routed distance between an agent's location and all possible destinations.
For this reason,
it is helpful to precompute the distances between all (or the most important) cells to all cells.
We call this step the routing matrix precomputation.
Precomputation simplifies the usage of the ShortestPathTree API of GraphHopper, as well as storing and reusing the results.

The runtime for precomputation of the routing matrix is depicted in \Cref{tab:Runtime}.
These runtimes depend on the size of the model area and, therefore, increase with the buffer radius.
For large model areas the precomputation of the routing matrix requires significant time.
In these cases, it is possible to reduce the spatial resolution of the routing grid (see \Cref{sec:Grid}) or
switch the distance metric to the Euclidean Distance.
Switching to the Euclidean Distance completely removes the need for precomputation,
but doing so is only advisable for test runs, as it skews the results significantly.

The runtime of the main simulation increases linearly with the number of agents and days.
\Cref{tab:Runtime} depicts the runtime of this component for one day and
1,000 to 1 million agents.

\section{Conclusion}
\label{sec:Conclusion}

In this paper,
we introduced the open-source mobility demand generator \ac{OMOD}.
\ac{OMOD} determines activity schedules for a population of agents for a user-specified area of interest.
These schedules state \textit{what} an agent does, \textit{where}, and for \textit{how long}.
The \textit{what} and \textit{how long} are sampled from probability distributions
calibrated with household travel survey data.
The \textit{where} is determined with a disaggregated destination choice model inspired by the gravity model concept.

\ac{OMOD} can be used in many different research fields,
like communications research,
energy system modeling,
epidemiology,
or for prototyping in transportation studies.
For example,
we use it in a publicly funded project \cite{ModelProjectESMRegio}
to determine the benefit of intelligent electric vehicle charging for operators of distribution grids.

We compare the generated mobility demand of \ac{OMOD} to the results of the German household travel survey \ac{MiD}.
This validation led to the following conclusions:

\begin{itemize}
    \item The trip destination distribution is satisfyingly reproduced for spatial resolutions up to \SI{1}{\kilo\meter}.
    \item Origin-destination matrices are significantly harder to reproduce.
          The results are satisfactory up to a resolution of \SI{5}{\kilo\meter}.
    \item The average daily driven distance error is negligible if the modeled area is large enough.
          For the best results we recommend an area of around \SI{10000}{\kilo\meter}.
    \item The share of agents conducting a specific activity at a each point in time matches the survey closely.
          The exception is the number of people currently moving.
          Here, a more sophisticated assignment process is necessary
          than the all-or-nothing approach used in our validation.
\end{itemize}

The validation uncovered several aspects in which the model could be improved.
These include:

\begin{itemize}
    \item The introduction of more explanatory variables.
          Especially spatially resolved socio-demographic features.
    \item The inclusion of non-linear effects of OpenStreetMap features on a location's attractiveness.
    \item The inclusion of mode choice, public transportation, and congestion effects in the destination choice step.
    \item The inclusion of household interactions.
\end{itemize}

Including these aspects will likely necessitate the inclusion of more input data sources,
such as more detailed census information and public transport schedules.

The biggest open question regards \ac{OMOD}'s ability
to translate to countries other than Germany.
Technically, \ac{OMOD} can be executed with a focus area that can be anywhere on Earth.
However, the current parameterization is calibrated with German household travel survey data
and has not yet been validated for other parts of the world.
In future work,
we aim to acquire mobility data for many more regions
and will use it to evaluate and improve \ac{OMOD}'s performance in as many places as possible.

\section*{Open-Source}

\ac{OMOD} is written in Kotlin (a modern JVM language) and is available on GitHub https://github.com/L-Strobel/omod under the MIT license.
To execute the model, the user only needs to
have Java installed on their machine,
download an OpenStreetMap file of the area they are interested in,
and define the focus area as a GeoJSON (for example, with https://geojson.io).
See the GitHub page for a step-by-step description of how to run the model.

\section*{Acknowledgment}
This model is created as part of the ESM-Regio project (https://www.bayern-innovativ.de/de/seite/esm-regio-en)
and is made possible through funding from the German Federal Ministry for Economic Affairs and Climate Action.

\appendix

\section{Discrete Distance Destination Choice Function}
\label{sec:smplDestChoice}

In this section,
we will explain how we reformulated the maximum likelihood problem described in \Cref{sec:deterrence}
to reduce the memory costs by digitizing the distances into the set $D$ of discrete bins.

The original maximum log-likelihood problem is:

\begin{equation}
    \begin{aligned}
        \underset{\theta}{arg\ max}\quad\quad &  \sum_{\forall (o, t) \in O} ln(P(o, t; \theta))
    \end{aligned}
    \label{eq:LogLike}
\end{equation}

where $O$ is the set of all observed origin-destination pairs in the \ac{MiD} data
and $P((o, t); \theta)$ is the probability that the building $t$ is the destination of a trip starting at $o$:

\begin{equation}
    P((o, t); \theta) = \frac{e^{\ln(A_o) + \ln(f(d_{o, t}; \theta))}}{\sum\limits_{\forall j \in Buildings}{e^{\ln(A_j) + \ln(f(d_{o, j}; \theta))}}}
    \label{eq:POptDetOrg}
\end{equation}

If we reformulate \Cref{eq:LogLike}, we get:

\begin{equation}
    \sum_{\forall (o, t) \in O} ln(A_o) + ln(f(d_{o, t}; \theta)) - ln(\sum\limits_{\forall j \in Buildings}{A_j e^{\ln(f(d_{o, j}; \theta))})}
    \label{eq:LogLikeSimpler}
\end{equation}

Ignoring constant terms and introducing $B_{o, d}$ for the set of buildings that have the distance $d$ to the origin,
we get:

\begin{equation}
    \sum_{\forall (o, t) \in O} ln(f(d_{o, t}; \theta)) - ln(\sum\limits_{\forall d \in D}{(e^{\ln(f(d; \theta))} \sum\limits_{\forall j \in B_{o, d}}{A_j})})
    \label{eq:LogLikeEvenSimpler}
\end{equation}

Note, that the term $\sum\limits_{\forall j \in B_{o, d}}{A_j}$ does not depend on optimization variables $\theta$.
Therefore, we can precalculate the term for all distances and observed origins before running the optimization.
If we digitize the distances in \SI{50}{\meter} wide bins, $D$ contains around \num{2e4} bins.
Therefore, with the \num{e5} observations,
we have to precompute and store \num{2e9} values, significantly less then the \num{9e10} values initially necessary for the distance matrix (with other simplifications already applied).
The memory consumption of the other terms in \Cref{eq:LogLikeEvenSimpler} is negligible.
Therefore, this approach enables us to precompute and store all necessary information
for the deterrence function parameterization in memory,
making the fit on all of Germany possible.

\section{Zonal Trip Attraction Metrics}
\label{sec:appendixAttraction}

This section contains all the quantitive metrics
that describe the similarity between the trip destination probability distributions
in the \ac{MiD} and that produced by \ac{OMOD}.

\begin{table}[H]
    \centering
    \begin{tabular}{llrrr}
        \toprule
        Activity & Resolution &  $R^2$ &  MAE [\%] &  Jensen-Shannon \\
        \midrule
              All &      \SI{5}{\kilo\meter} &  0.935 &                1.692 &           0.073 \\
              All &      \SI{1}{\kilo\meter} &  0.769 &                0.289 &           0.210 \\
              All &         \SI{500}{\meter} &  0.215 &                0.162 &           0.407 \\
    \textit{home} &      \SI{5}{\kilo\meter} &  0.849 &                2.472 &           0.139 \\
    \textit{home} &      \SI{1}{\kilo\meter} &  0.728 &                0.341 &           0.236 \\
    \textit{home} &         \SI{500}{\meter} &  0.513 &                0.171 &           0.408 \\
    \textit{work} &      \SI{5}{\kilo\meter} &  0.993 &                0.859 &           0.073 \\
    \textit{work} &      \SI{1}{\kilo\meter} &  0.646 &                0.479 &           0.326 \\
    \textit{work} &         \SI{500}{\meter} &  0.114 &                0.225 &           0.532 \\
\textit{shopping} &      \SI{5}{\kilo\meter} &  0.820 &                2.673 &           0.142 \\
\textit{shopping} &      \SI{1}{\kilo\meter} &  0.565 &                0.478 &           0.333 \\
\textit{shopping} &         \SI{500}{\meter} & -0.054 &                0.218 &           0.529 \\
   \textit{other} &      \SI{5}{\kilo\meter} &  0.899 &                2.039 &           0.102 \\
   \textit{other} &      \SI{1}{\kilo\meter} &  0.644 &                0.430 &           0.286 \\
   \textit{other} &         \SI{500}{\meter} &  0.008 &                0.208 &           0.485 \\
  \textit{school} &      \SI{5}{\kilo\meter} &  0.950 &                2.465 &           0.168 \\
  \textit{school} &      \SI{1}{\kilo\meter} &  0.764 &                0.470 &           0.370 \\
  \textit{school} &         \SI{500}{\meter} &  0.264 &                0.231 &           0.522 \\
        \bottomrule
    \end{tabular}
    \caption{Kassel: trip attraction metrics}
\end{table}

\begin{table}[H]
    \centering
    \begin{tabular}{llrrr}
        \toprule
        Activity & Resolution & $R^2$ &  MAE [\%] &  Jensen-Shannon \\
        \midrule
              All &      \SI{5}{\kilo\meter} & 0.874 &                0.739 &           0.090 \\
              All &      \SI{1}{\kilo\meter} & 0.641 &                0.104 &           0.215 \\
              All &         \SI{500}{\meter} & 0.290 &                0.051 &           0.388 \\
    \textit{home} &      \SI{5}{\kilo\meter} & 0.927 &                0.745 &           0.092 \\
    \textit{home} &      \SI{1}{\kilo\meter} & 0.796 &                0.100 &           0.215 \\
    \textit{home} &         \SI{500}{\meter} & 0.576 &                0.053 &           0.398 \\
    \textit{work} &      \SI{5}{\kilo\meter} & 0.909 &                1.064 &           0.134 \\
    \textit{work} &      \SI{1}{\kilo\meter} & 0.524 &                0.167 &           0.333 \\
    \textit{work} &         \SI{500}{\meter} & 0.169 &                0.068 &           0.490 \\
\textit{shopping} &      \SI{5}{\kilo\meter} & 0.792 &                1.125 &           0.148 \\
\textit{shopping} &      \SI{1}{\kilo\meter} & 0.299 &                0.210 &           0.372 \\
\textit{shopping} &         \SI{500}{\meter} & 0.052 &                0.078 &           0.535 \\
   \textit{other} &      \SI{5}{\kilo\meter} & 0.862 &                0.937 &           0.122 \\
   \textit{other} &      \SI{1}{\kilo\meter} & 0.529 &                0.145 &           0.291 \\
   \textit{other} &         \SI{500}{\meter} & 0.145 &                0.062 &           0.447 \\
  \textit{school} &      \SI{5}{\kilo\meter} & 0.866 &                1.175 &           0.204 \\
  \textit{school} &      \SI{1}{\kilo\meter} & 0.533 &                0.209 &           0.421 \\
  \textit{school} &         \SI{500}{\meter} & 0.142 &                0.089 &           0.590 \\
        \bottomrule
    \end{tabular}    
    \caption{Nuremberg: trip attraction metrics}
\end{table}

\begin{table}[H]
    \centering
    \begin{tabular}{llrrr}
        \toprule
        Activity & Resolution & $R^2$ &  MAE [\%] &  Jensen-Shannon \\
        \midrule
              All &      \SI{5}{\kilo\meter} & 0.950 &                0.390 &           0.089 \\
              All &      \SI{1}{\kilo\meter} & 0.613 &                0.047 &           0.206 \\
              All &         \SI{500}{\meter} & 0.359 &                0.023 &           0.363 \\
    \textit{home} &      \SI{5}{\kilo\meter} & 0.907 &                0.454 &           0.101 \\
    \textit{home} &      \SI{1}{\kilo\meter} & 0.765 &                0.052 &           0.216 \\
    \textit{home} &         \SI{500}{\meter} & 0.572 &                0.023 &           0.361 \\
    \textit{work} &      \SI{5}{\kilo\meter} & 0.873 &                0.649 &           0.146 \\
    \textit{work} &      \SI{1}{\kilo\meter} & 0.562 &                0.087 &           0.359 \\
    \textit{work} &         \SI{500}{\meter} & 0.309 &                0.034 &           0.513 \\
\textit{shopping} &      \SI{5}{\kilo\meter} & 0.492 &                0.876 &           0.186 \\
\textit{shopping} &      \SI{1}{\kilo\meter} & 0.390 &                0.090 &           0.358 \\
\textit{shopping} &         \SI{500}{\meter} & 0.202 &                0.032 &           0.486 \\
   \textit{other} &      \SI{5}{\kilo\meter} & 0.919 &                0.531 &           0.126 \\
   \textit{other} &      \SI{1}{\kilo\meter} & 0.458 &                0.065 &           0.274 \\
   \textit{other} &         \SI{500}{\meter} & 0.128 &                0.029 &           0.447 \\
  \textit{school} &      \SI{5}{\kilo\meter} & 0.771 &                0.793 &           0.200 \\
  \textit{school} &      \SI{1}{\kilo\meter} & 0.381 &                0.113 &           0.473 \\
  \textit{school} &         \SI{500}{\meter} & 0.153 &                0.044 &           0.632 \\
        \bottomrule
    \end{tabular}
    \caption{Hamburg: trip attraction metrics}
\end{table}

\begin{acronym}[ABCDEFGHIJK]
	\acro{OMOD}{OpenStreetMap \allowbreak Mobility \allowbreak Demand \allowbreak Generator \allowbreak}
	\acro{OSM}{OpenStreetMap}
    \acro{POI}{point of interest}
    \acro{MiD}{Mobilität-in-Deutschland 2017}
    \acro{MNL}{multinomial logit model}
\end{acronym}

\bibliographystyle{unsrt}
\bibliography{omod.bib}

\end{document}